\documentclass[prd,nofootinbib,superscriptaddress,notitlepage]{revtex4-1}
\usepackage[a4paper,left=1.5cm,right=1.5cm,top=3cm,bottom=3cm]{geometry}

\usepackage{tensor}
\usepackage{xcolor}
\usepackage{enumerate}
\usepackage{framed}
\usepackage{mdframed}
\usepackage{amsmath}
\usepackage{amsfonts}
\usepackage{mathrsfs}
\usepackage{amssymb}
\usepackage{graphicx, rotating}
\usepackage{epsfig}
\usepackage{latexsym}
\usepackage{graphicx}
\usepackage{color}
\usepackage{amsmath,bm,amssymb}
\usepackage{slashed}
\usepackage{hyperref}
\usepackage[font=footnotesize,labelfont=]{caption}
\hypersetup{colorlinks=true, citecolor=bluscuro, linkcolor=black, urlcolor=bluscuro}
\definecolor{rossos}{cmyk}{0,1,1,0.55}
\definecolor{bluscuro}{rgb}{0.15, 0.2, .85}
\definecolor{bluchiaro}{cmyk}{1,.3,0.,0.1}
\usepackage[T1]{fontenc} 

\usepackage{tabularx}

\usepackage{subcaption}
\usepackage{ragged2e}
\DeclareCaptionJustification{justified}{\justifying}
\usepackage{tcolorbox}

\def\0{\vec{0}}

\newcommand{\bea}{\begin{eqnarray}}
	\newcommand{\eea}{\end{eqnarray}}
\def\beq{\begin{equation}}
	\def\eeq{\end{equation}}

\def\d{{\rm d}}

\def\beqa{\begin{eqnarray}}

	\def\eeqa{\end{eqnarray}}
\def\lsim{\mathrel{\rlap{\lower4pt\hbox{\hskip0.5pt$\sim$}}
		\raisE_1pt\hbox{$<$}}}         
\def\gsim{\mathrel{\rlap{\lower4pt\hbox{\hskip0.5pt$\sim$}}
		\raisE_1pt\hbox{$>$}}}         

\def\d{{\rm d}}

\def\d{{\rm d}}

\def\PBH{\text{\tiny PBH}}

\def\eeqa{\end{eqnarray}}
\numberwithin{equation}{section}

\def\bq{\begin{quote}}
\def\eq{\end{quote}}

 at 10truept

\newcommand{\llp}{\left [}
\newcommand{\rrp}{\right ]}

\newcommand{\lp}{\left (}
\newcommand{\rp}{\right )}

\newcommand{\be}{\begin{equation}\begin{aligned}}
	\newcommand{\ee}{\end{aligned}\end{equation}}

\usepackage[normalem]{ulem}

\def\cs2{c_{\rm{s}}^2}

\def\U0{{\bar U_0}}
\def\bi{\begin{itemize}}
\def\ei{\end{itemize}}

\def\U{{\cal{U}}}

\begin{document}
\title{
The Minimum Testable Abundance of Primordial Black Holes\\
at Future Gravitational-Wave Detectors
}

\author{Valerio De Luca}
\address{D\'epartement de Physique Th\'eorique and Centre for Astroparticle Physics (CAP), Universit\'e de Gen\`eve, 24 quai E. Ansermet, CH-1211 Geneva, Switzerland}
\address{Dipartimento di Fisica, Sapienza Università 
di Roma, Piazzale Aldo Moro 5, 00185, Roma, Italy}

\author{Gabriele Franciolini}
\address{D\'epartement de Physique Th\'eorique and Centre for Astroparticle Physics (CAP), Universit\'e de Gen\`eve, 24 quai E. Ansermet, CH-1211 Geneva, Switzerland}

\author{Paolo~Pani}
\affiliation{Dipartimento di Fisica, Sapienza Università 
di Roma, Piazzale Aldo Moro 5, 00185, Roma, Italy}
\affiliation{INFN, Sezione di Roma, Piazzale Aldo Moro 2, 00185, Roma, Italy,}

\author{Antonio~Riotto}
\address{D\'epartement de Physique Th\'eorique and Centre for Astroparticle Physics (CAP), Universit\'e de Gen\`eve, 24 quai E. Ansermet, CH-1211 Geneva, Switzerland}

\date{\today}

\begin{abstract}
\noindent
The next generation of gravitational-wave experiments, such as Einstein Telescope, Cosmic Explorer and LISA, will test the primordial black hole scenario. We provide a forecast for the minimum testable value of the abundance of primordial black holes as a function of their masses for both the unclustered and clustered spatial distributions at formation. In particular, we show that these instruments may test abundances, relative to the dark matter, as low as  $10^{-10}$.

\end{abstract}

\maketitle

\section{Introduction}
\renewcommand{\theequation}{1.\arabic{equation}}
\setcounter{equation}{0}
\label{intro}

\noindent
The recent flow of measurements detected by the LIGO/Virgo collaboration~\cite{Abbott:2020niy} from   gravitational wave (GW) signals  produced by Black Hole (BH) mergers has revived the possibility that Primordial BHs (PBHs) may comprise a significant fraction of the dark matter in the universe~\cite{Bird:2016dcv,Clesse:2016vqa,Sasaki:2016jop,Eroshenko:2016hmn, Wang:2016ana,
Ali-Haimoud:2017rtz,Clesse:2017bsw, Chen:2018czv,Raidal:2018bbj,Garriga:2019vqu, Hutsi:2019hlw, Vaskonen:2019jpv, Gow:2019pok,Wu:2020drm,DeLuca:2020bjf, Hall:2020daa,DeLuca:2020sae,Wong:2020yig,Hutsi:2020sol,Kritos:2020wcl,DeLuca:2021wjr,Deng:2021ezy,Kimura:2021sqz}. 

In particular, current data allow for the possible presence of a considerable fraction of merger events ascribable to PBHs~\cite{Franciolini:2021tla} as long as their abundance $f_\PBH = \Omega_\PBH/\Omega_\text{\tiny DM}$ in units  of the dark matter one  is below $10^{-3}$ in the LIGO/Virgo mass range. Whether or not Third Generation~(3G) ground-based detectors, such as the Einstein Telescope (ET)~\cite{Hild:2010id}, the Cosmic Explorer (CE)~\cite{Reitze:2019iox}, and the future Laser Interferometer Space Antenna~(LISA)~\cite{amaroseoane2017laser}, will confirm this hypothesis, 
even a smaller value of $f_\PBH$ may still have an impact on the cosmological evolution as far as large scale structure is concerned (see, for example, Refs.~\cite{Adamek:2019gns,Inman:2019wvr, Bertone:2019vsk, Boldrini:2019isx, DeLuca:2020jug}).  
It is therefore timely and interesting to ask what is the minimum testable value of $f_\PBH$ by future GW experiments, a question we address in this paper. 

For clarity's sake, we state our starting assumptions in the following. We first  consider the standard PBH scenario where the PBH binaries form in the early universe \cite{Nakamura:1997sm,Ioka:1998nz} from an initially Poisson distributed population~\cite{Ali-Haimoud:2018dau,Desjacques:2018wuu,Ballesteros:2018swv,MoradinezhadDizgah:2019wjf,Inman:2019wvr,DeLuca:2020jug}. As the merger rate might be enhanced if there is  significant spatial clustering  at PBH formation (e.g. if a large-small scale correlation is present thanks to a primordial non-Gaussianity~\cite{Tada:2015noa, Raidal:2017mfl, Atal:2019erb, Young:2019gfc}), we also consider the impact of such clustering on the determination of the mimimum value of $f_\PBH$, without however including the clustering evolution and the potential suppression effects from binary interactions in local clusters (see e.g. \cite{Jedamzik:2020ypm}), which are currently poorly understood. Therefore, the clustered scenario should be regarded as corresponding to the maximum possible value of the merger rate. 
Secondly, our results refer to the case of a  monochromatic PBH mass function and we leave the study of the effects of accretion on the merger rate~\cite{DeLuca:2020qqa} and the comparison with constraints on $f_\PBH$ \cite{DeLuca:2020fpg} to further investigation due to the uncertainties of the treatment at large PBH masses. 

We will present results for both the space-based detector LISA and ground-based detectors like ET (we note that similar conclusions apply to the CE experiment). 

Finally, assuming the actual value of $f_\PBH$ is above the minimum detectability threshold, i.e. the corresponding PBH population is able to produce enough merger events to be seen by future GW detectors, the next natural question concerns the possibility to confidently identify the primordial nature of some GW events.
This can be achieved at least in two cases: i) if at least one of the binary components have a subsolar mass, since no other astrophysical compact object is expected in that range; and ii) if the mergers occur at sufficiently high redshift, where astrophysical sources do not contribute. We will then provide forecast for the minimal $f_\PBH$ required for these cases, thus allowing to confidently assign a fraction of the observed merger signals to PBHs.

The paper is structured as follows. In Section~II we characterise the PBH merger rate, in Section~III we provide a forecast for the minimum value of $f_\PBH$ necessary to be testable by future detectors. In Section~IV we discuss the minimum PBH abundance sufficient to obtain evidence of PBHs and, finally, in Section~V we draw our conclusions.

\section{The PBH merger rate}
\renewcommand{\theequation}{2.\arabic{equation}}
\setcounter{equation}{0}
\label{sec2}

\noindent
Before matter-radiation equality, PBH binaries may be assembled through random gravitational decoupling from the Hubble flow if their distance $x$ is smaller than the comoving distance~\cite{Nakamura:1997sm, Ioka:1998nz}
\begin{equation}
\tilde{x} = \lp \frac{3}{4\pi} \frac{m_1+m_2}{a_\text{\tiny eq}^3\rho_\text{\tiny eq}} \rp^{1/3},
\end{equation}
in terms of the scale factor $a_\text{\tiny eq}$ and energy density $\rho_\text{\tiny eq}$ at matter-radiation equality.
Assuming PBHs come from Gaussian curvature perturbations, they are expected to follow a  Poisson spatial distribution at formation~\cite{Ali-Haimoud:2018dau,Desjacques:2018wuu,Ballesteros:2018swv,MoradinezhadDizgah:2019wjf,Inman:2019wvr}. 
In this standard scenario,
the differential merger rate of BH binaries of primordial origin is given by~\cite{Raidal:2018bbj}
\begin{align}
\frac{\d^2 R_\text{\tiny PBH} }{\d m_1 \d m_2}= 
\frac{1.6 \times 10^6}{\rm Gpc^3 \, yr}
f_\PBH^{\frac{53}{37}} \,
\lp \frac{t}{t_0} \rp^{-\frac{34}{37}}  
 \lp \frac{M_\text{\tiny tot}}{M_\odot} \rp^{-\frac{32}{37}}  
 \llp \frac{m_1 m_2}{(m_1+m_2)^2}\rrp^{-\frac{34}{37}} 
S\lp M_\text{\tiny tot}, f_\PBH,\psi  \rp
\psi(m_1) \psi (m_2),
\label{PBHrate}
\end{align}
which is defined in terms of the current age of the universe $t_0$, the
total mass of the binary $M_\text{\tiny tot}= m_1+m_2$ and the PBH mass function $\psi (m)$ at formation.
The suppression factor $S\lp M_\text{\tiny tot}, f_\PBH,\psi  \rp \equiv S_1 \times S_2$
takes into account the reduction of the PBH merger rate arising from interactions with surrounding inhomogeneities in the dark matter fluid and neighbouring PBHs~\cite{Ali-Haimoud:2017rtz,Raidal:2018bbj} around the binary formation epoch, while the second term $S_2$ tracks the successive disruption of binaries in clusters or sub-structure environments \cite{Vaskonen:2019jpv,Jedamzik:2020ypm,Young:2020scc,Jedamzik:2020omx,DeLuca:2020jug,Tkachev:2020uin,Hutsi:2020sol}, given that PBH binaries are expected to form before the matter-radiation equality. The two pieces read~\cite{Hutsi:2020sol}
\begin{align}
	S_1 (M_\text{\tiny tot}, f_\PBH, \psi)& \thickapprox 1.42 \llp \frac{\langle m^2 \rangle/\langle m\rangle^2}{\bar N(y) +C} + \frac{\sigma ^2_\text{\tiny M}}{f^2_\PBH}\rrp ^{-21/74} \exp \llp -  \bar N(y ) \rrp 
	\qquad 
	\text{with}
		\qquad 
		\bar N(y) \equiv \frac{M_\text{\tiny tot}}{\langle m \rangle } \frac{f_\PBH}{f_\PBH+ \sigma_\text{\tiny M}},
		\label{S1}
	\\
	S_2 (x) & \thickapprox \text{min} \llp 1, 9.6 \cdot 10^{-3} x ^{-0.65} \exp \lp 0.03 \ln^2 x \rp  \rrp 
	\qquad 	\qquad 	\,\,\ 
	\text{with}
		\qquad 
		x \equiv (t(z)/t_0)^{0.44} f_\PBH,
\end{align}
in terms of the rescaled variance of the matter density perturbations $\sigma_\text{\tiny M}^2 \simeq 3.6 \cdot 10^{-5}$. For a monochromatic mass function peaked at the mass scale $m_\PBH$, the expectation values simplify to become $\langle m \rangle = \langle m^2 \rangle^{1/2} = m_\PBH$, and $S_1$ is independent on the PBH mass. The suppression due to disruption in PBH sub-structures $S_2$ is negligible for small enough values of PBH abundance, i.e. $f_\PBH \lesssim 0.003$, compatible with the results obtained with cosmological N-body simulations~\cite{Inman:2019wvr}.
In this case, the merger rate evolution with redshift is dictated by the relation $R_\PBH \approx (t/t_0)^{-34/37}$.

We remark that for initially Poisson distributed PBHs
the merger rate of binaries produced via gravitational capture in the late-time universe is subdominant with respect to the early-universe one \cite{Ali-Haimoud:2017rtz,Raidal:2017mfl,Korol:2019jud,Vaskonen:2019jpv,DeLuca:2020jug}.

\subsection{Maximum merger rate of clustered PBHs}
In order to find the minimum PBH abundance testable by future detectors, in this subsection we consider the largest possible PBH merger rate coming from primordial binaries. 
The formation of binaries in the early universe can be enhanced if PBHs are clustered at formation. By maximizing the impact of initial clustering and neglecting the potential successive dynamical suppression of the merger rate due to binary disruption in sub-structures, one can estimate the maximum $R_\text{\tiny PBH}^\text{\tiny max}$.

If PBHs are generated from non-Gaussian curvature perturbations, which correlate short and long scales, their spatial distribution at formation differs from the Poisson one, and PBHs can be clustered at formation. This will enhance the formation of binaries in the very early universe. However, for a large enough PBH abundance and strong clustering, the typical semi-major axis of a PBH binary becomes so small that the coalescence of the binaries may happen much before the present epoch, eventually suppressing the corresponding present merger rate.

\begin{figure}[t!]
	\centering
	\includegraphics[width=0.7 \linewidth]{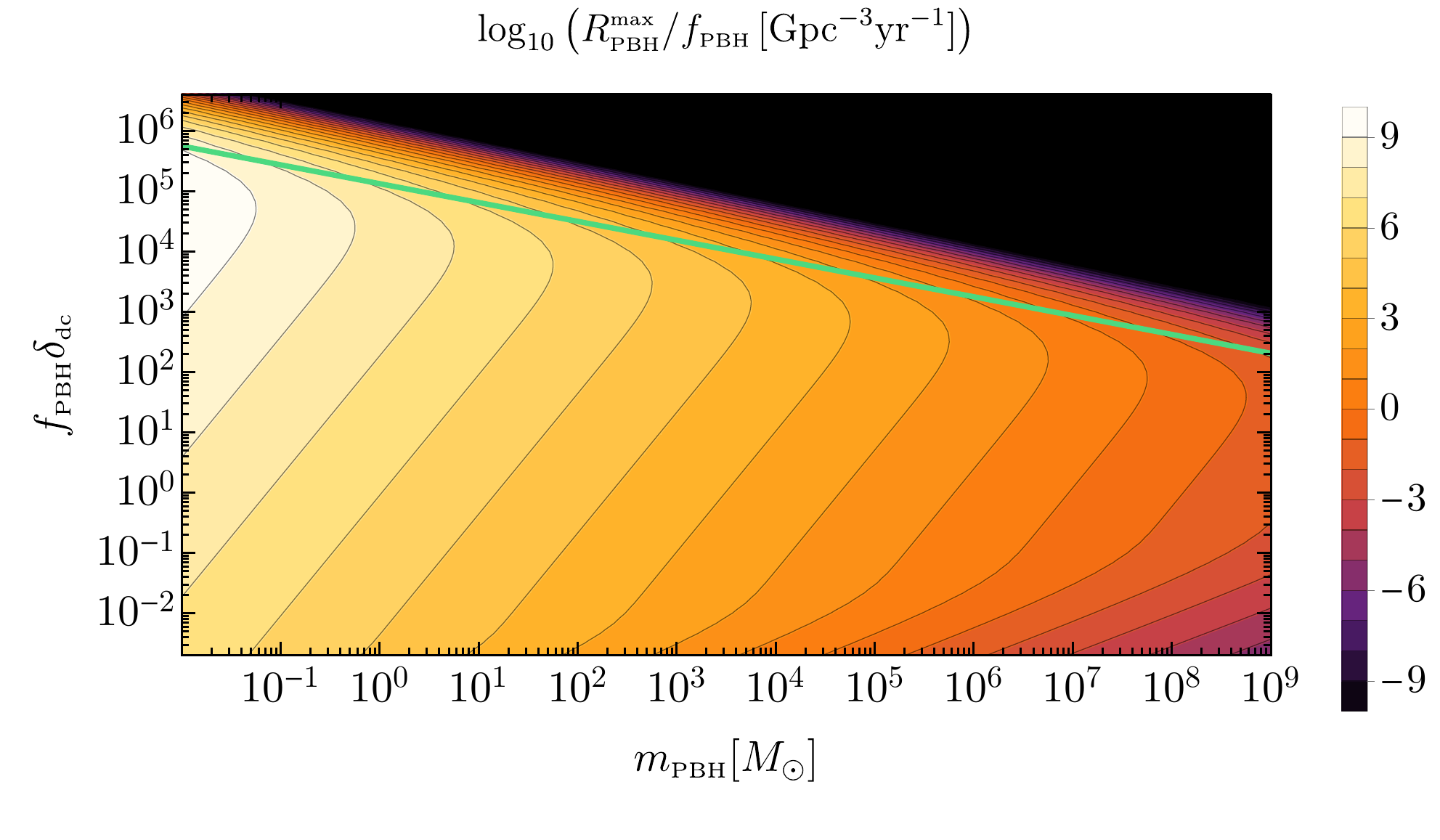}
	\caption{\it Contour plot of the maximum present merger rate of primordial binaries in terms of the PBH mass and the local abundance $f_\text{\tiny \rm PBH} \delta_{\rm dc}$. Above the green line the exponential suppression takes over, limiting the merger rate.
In practice, the maximum merger rate scales overall as $(t(z)/t_0)^{-1}$ in the interesting redshift range $z\lesssim 50$.}
	\label{fig: 1}
\end{figure}

Following Ref.~\cite{Raidal:2017mfl}, one can assume a PBH correlation function $\xi_\PBH$ which is constant up to the binary scale $\tilde x$ at the decoupling epoch as
\begin{equation}
1 + \xi_\PBH(x) \approx \delta_\text{\tiny dc}\,, \quad {\rm if }\quad x < \tilde{x},
\end{equation}
as obtained in simple non-Gaussian models of PBH formation \cite{Tada:2015noa,Suyama:2019cst,Atal:2020igj,DeLuca:2021hcf}.
The upper bound on the PBH merger rate, i.e  not accounting for the dynamical suppression due to the binary disruption in sub-structures, for a monochromatic population of PBHs with mass $m_\PBH$ is then found to be~\cite{Raidal:2017mfl}
\begin{align}\label{rate_clusteredPBH}
R^\text{\tiny max}_\PBH
	=& \frac{6.2 \cdot 10^{4}}{{\rm Gpc}^3 {\rm yr}} 
	\delta_\text{\tiny dc}^{16/37} 
	f_\PBH^{53/37}
	\lp \frac{t}{t_0}\rp^{-34/37}  
	\lp \frac{m_\PBH}{30 M_\odot}\rp^{-32/37}
\nonumber \\
\times &\left\{ \Gamma\left[\frac{58}{37}, 
9.5 \cdot 10^{-5} \delta_\text{\tiny dc} f_\PBH \lp\frac{m_\PBH}{30 M_\odot} \rp^{5/16}
\lp \frac{t}{t_0} \rp^{3/16}
\right] - 
\Gamma\left[ \frac{58}{37},
850 \delta_\text{\tiny dc} f_\PBH \lp\frac{m_\PBH}{30 M_\odot} \rp^{-5/21}
\lp \frac{t}{t_0} \rp^{-1/7}
\right] \right\}.
\end{align}
In Fig.~\ref{fig: 1} we show the contour plot of the maximum PBH merger rate of clustered PBHs in terms of their local abundance and mass. As one can appreciate, the merger rate is found to increase for small masses and large values of the PBH correlation function up to the solid green line, determined by the critical value  where the exponential suppression takes over and drastically reduce the merger rate.

To gain further indications on the theoretical maximum merger rate, let us investigate Eq.~\eqref{rate_clusteredPBH} in the highly clustered scenario.
One can expand the merger rate for large values of the local abundance $\delta_\text{\tiny dc} f_\PBH \gg 1$ as
\begin{align}
\label{PBHrateclu}
R^\text{\tiny max}_\PBH
	&\simeq \frac{1.9\times 10^6}{{\rm Gpc}^3 {\rm yr}} f_\PBH 	\lp \frac{t}{t_0}\rp^{-1}  \lp \frac{m_\PBH}{30 M_\odot}\rp^{-1}  \lp 1 + 1.7 \cdot 10^{-4} \Delta_\text{\tiny dc} \rp \exp\llp- 9.5 \cdot 10^{-5} \Delta_\text{\tiny dc}\rrp,
\end{align}
where we defined 
\be
\label{critdc}
\Delta_\text{\tiny dc} = \delta_\text{\tiny dc} f_\PBH 	\lp \frac{t}{t_0}\rp^{3/16}  \lp \frac{m_\PBH}{30 M_\odot}\rp^{5/16}.
\ee
Let us focus on the enhancement given by the local abundance $\delta_\text{\tiny dc} f_\PBH$, i.e. the last two terms in Eq.~(\ref{PBHrateclu}). This expansion shows that large values of $\delta_\text{\tiny dc}$ lead to an enhancement of the rate, up until the exponential suppression becomes dominant. 
For fixed values of mass $m_\PBH$ and overall abundance $f_\PBH$, the exponential suppression is preferentially active at low redshift (i.e. $\Delta_\text{\tiny dc}  \propto \lp {t}/{t_0}\rp^{3/16}$), as most of the binaries are produced with small separation and merger times. 
This means that the theoretical maximum merger rate before the exponential suppression takes over can be estimated requiring the combination of parameters to approach the value $\Delta_\text{\tiny dc} \approx 4.6 \cdot 10^3$, obtained by maximising Eq.~\eqref{PBHrateclu}. 

In the following we will search for the optimal value of $\delta_\text{\tiny dc}$ leading to the maximum number of events at GW detectors, in order to asses the minimum testable PBH abundance.  
This can be found by determining the critical local abundance for which the merger rate for clustered PBHs is maximum at redshift $z=0$. Due to the weak dependence on redshfit of the exponential suppression in the interesting parameter space ($z\lesssim 50$), a similar result can be obtained by requiring the maximum number of events integrated in the observable redshift window.
It is also interesting to notice that in the case of strongly clustered PBHs, the overall merger rate evolution with redshift is modified and follows the behaviour $R_\PBH^\text{\tiny max} \approx (t/t_0)^{-1}$.

These considerations, however, do not account for the further potential suppression of the merger rate due to interactions in dense environments.
As such, the forecast for the minimum value of $f_\PBH$ in the case of clustered distribution has to be considered as the 
most optimistic case, i.e. accounting for a potential suppression would increase the minimum value of $f_\PBH$.
We also remark that in this extreme scenario, we do not expect our results to be changed including the late time merger rate, that is the one due to PBH binaries formed at very low redshifts through GW captures.
This is because, for the relevant small values of $f_\PBH$, the development of large scale structures dominated by a different dark matter component is not supposed to 
be affected by PBHs and to enhance the probability of PBH encounters in the late-time universe.
\begin{figure}[t!]
	\centering
	\includegraphics[width=0.7 \linewidth]{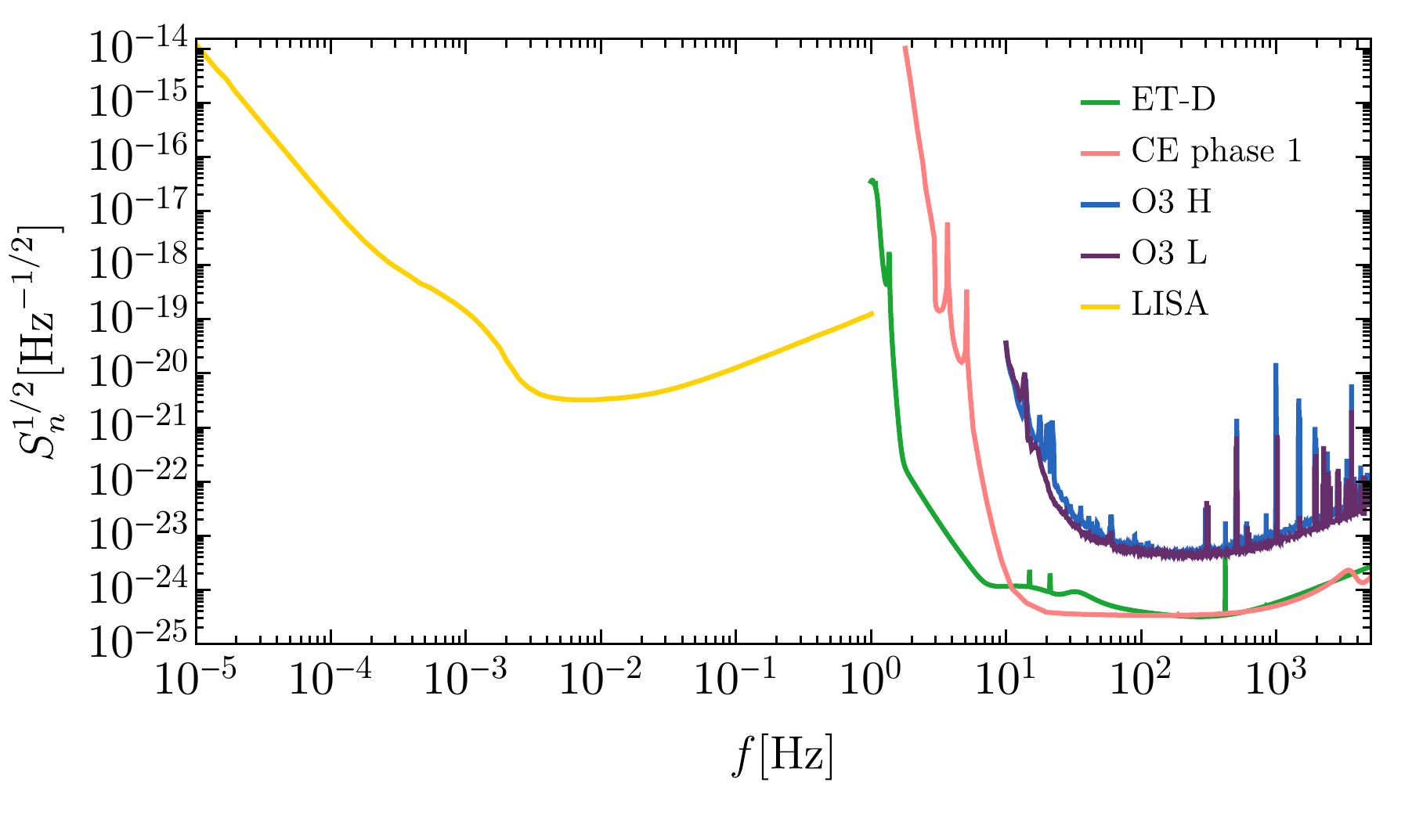}
	\caption{\it Noise curves for the LIGO Hanford (H) and Livingston (L) detector during the O3 run~\cite{Abbott:2020niy}, both the 3G detectors Einstein Telescope (ET) at design sensitivity from~\cite{Hild:2010id} and Cosmic Explorer (CE) during phase 1 from~\cite{ce}, and the LISA experiment~\cite{Kaiser:2020tlg}.}
	\label{psd}
\end{figure}

\section{Forecast for future Gravitational-Wave detectors}
\renewcommand{\theequation}{3.\arabic{equation}}
\setcounter{equation}{0}
\label{sec3}
\noindent
In this section we provide the forecast on the PBH abundance, both considering the event rate and the SGWB at 3G detectors like the ET and LISA.

\subsection{Event rate of resolvable mergers}
\noindent
From the PBH merger rate, one can compute the expected rate of merger events as
\be
N_\text{\tiny det} = \int \d z \d m_1 \d m_2 \frac{1}{1+z} \frac{\d V_c (z)}{\d z}  \frac{\d^2 R_\text{\tiny PBH} }{\d m_1 \d m_2} p_\text{\tiny det} (m_1, m_2, z),
\ee
for a given GW experiment, where the additional factor of redshift $1/(1+z)$ has been introduced to account for the difference in the clock rates at the time of merger and detection, while the comoving volume per unit redshift is provided in Ref.~\cite{Dominik:2014yma}.
The factor $p_\text{\tiny det}(m_1, m_2, z)$ accounts for the probability of detection of a binary, averaged over the source orientation, as a function of the signal-to-noise ratio (SNR). Many details on its computation are provided in Appendix A of Ref.~\cite{DeLuca:2021wjr}. 

In Fig~\ref{psd} we show the strain noise $S_n$ for the current O3 LIGO/Virgo observation runs, future 3G detectors like Einstein Telescope and Cosmic Explorer, and the one for the LISA experiment.
In particular, the binary detectability has been computed using the noise power spectral densities of the ET and LISA experiments which are found in Refs.~\cite{Hild:2010id} and \cite{Kaiser:2020tlg} respectively. Also, we adopt the non-precessing waveform model IMRPhenomD \cite{Husa:2015iqa,Khan:2015jqa}. 

\begin{figure}[t!]
	\centering
	\includegraphics[width=0.49 \linewidth]{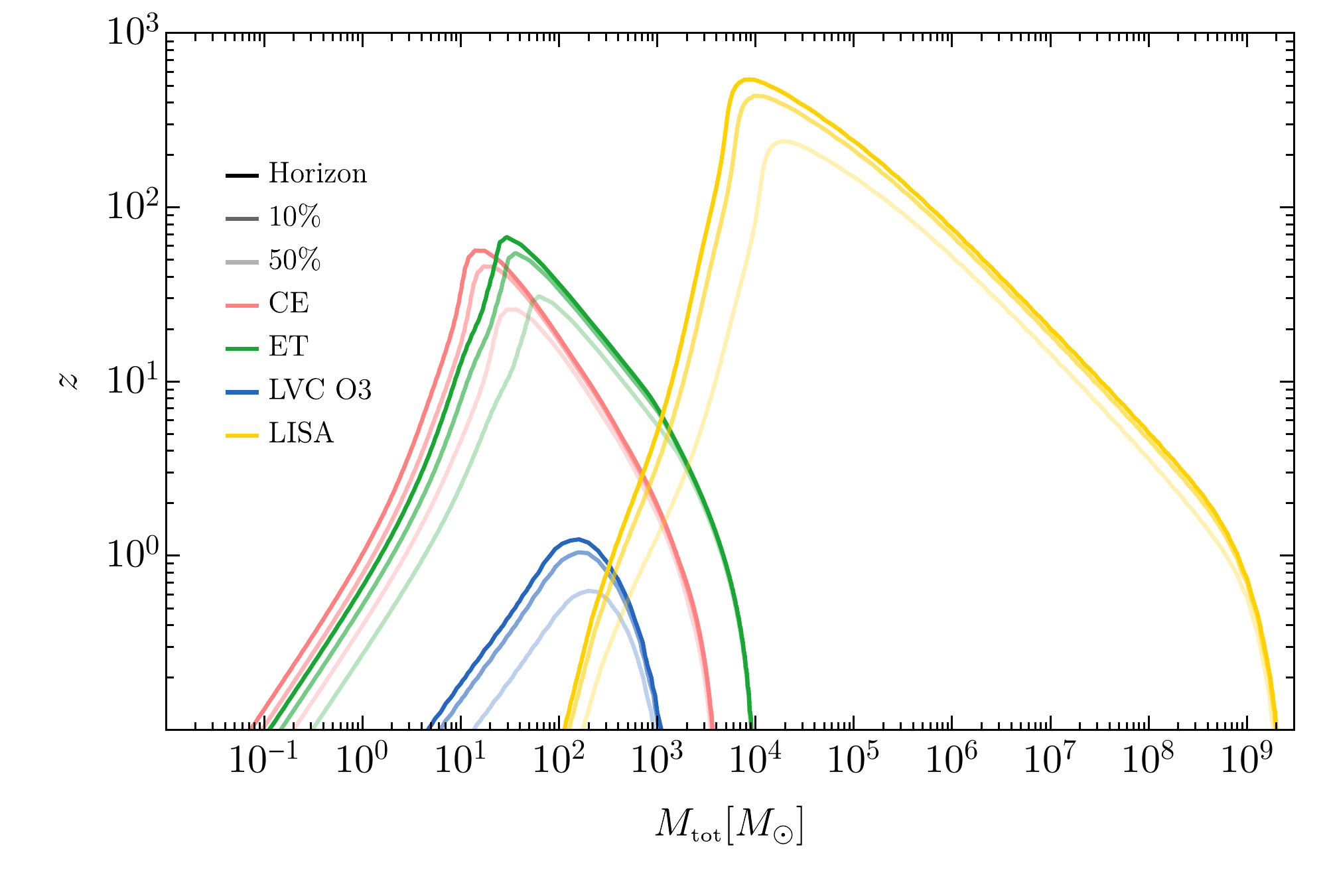}
	\includegraphics[width=0.49 \linewidth]{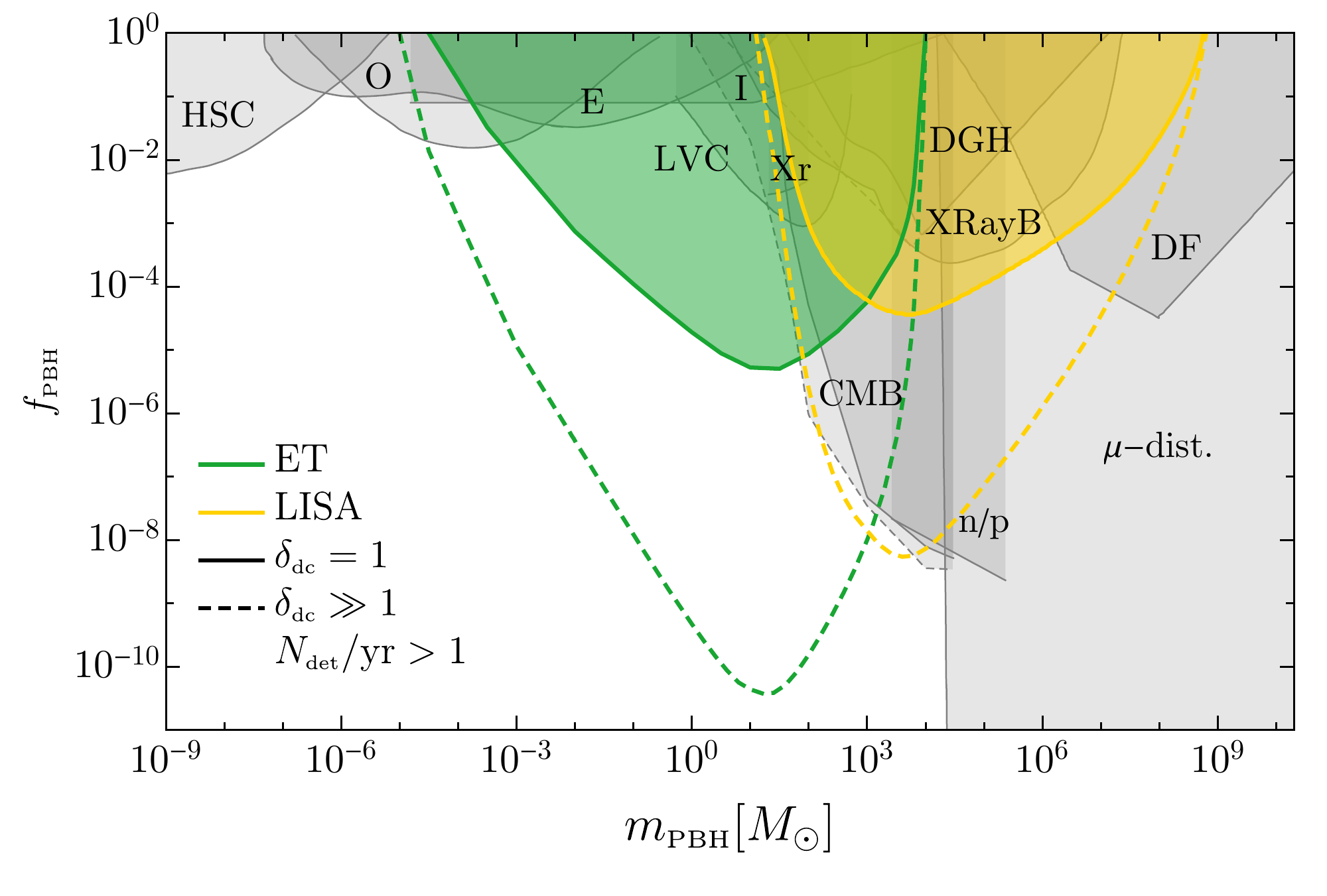}
	\caption{\it 
	{\bf Left:} Horizon redshift for  several GW experiments in terms of the total mass of a symmetric binary.
	In the plot we refer to the ``Horizon'' as the maximum distance at which a binary can in principle be observed if optimally oriented with respect to the detectors, while $10\%$,  $50\%$ as the redshift at which those fraction of binaries are observable.
 Those values correspond to values of ${\rm SNR}=\{ 8,10,19 \}$, respectively.
	{\bf Right:} Minimum PBH abundance required to have at least one event per year at the ET (green) and LISA (yellow) experiments, for both cases of Poisson distributed (solid) and strongly clustered (dashed) PBHs. Superimposed on these curves we show the present observational constraints (gray) coming from microlensing searches by Subaru HSC~\cite{Niikura:2017zjd, Smyth:2019whb}, MACHO/EROS (E)~\cite{Alcock:2000kd, Allsman:2000kg}, Ogle (O)~\cite{Niikura:2019kqi} and Icarus (I)~\cite{Oguri:2017ock}, X-rays (Xr)~\cite{Manshanden:2018tze} and X-Ray binaries (XRayB)~\cite{Inoue:2017csr}, CMB anisotropies~\cite{Serpico:2020ehh}, Dwarf Galaxy heating (DGH)~\cite{Lu:2020bmd,Takhistov:2021aqx}, dynamical friction (DF)~\cite{Carr:2018rid}, the neutron-to-proton ratio (n/p)~\cite{Inomata:2016uip} and CMB $\mu$-distortions~\cite{Nakama:2017xvq}.}
	\label{fig: 3}
\end{figure}

In the left panel of Fig.~\ref{fig: 3} we show the horizon redshift in terms of the total mass of a symmetric binary for several present and future GW experiments like ET and LISA. Requiring the detection of at least one event per year at those experiments, $N_\text{\tiny det} > 1/{\rm yr}$, one can determine a minimum value for the PBH abundance for both cases of Poisson distributed (Eq.~\eqref{PBHrate}) and clustered (Eq.~\eqref{PBHrateclu}) PBHs, where for the second case we have assumed the theoretical maximum merger rate before the exponential suppression takes over.

The result is shown in the right panel of Fig.~\ref{fig: 3} for a monochromatic population of PBHs, where the minimum testable PBH abundance has been superimposed with present observational constraints coming from several independent searches.
One can notice that the effect of clustering in enhancing the merger rate results in a much smaller value of the testable PBH abundance with respect to the Poisson case, although we remind that the case of clustering is optimistic, since it neglects a possible suppression of the merger rates.
Notice also that the constraints are strictly valid for a monochromatic PBH mass function and for an initial Poisson spatial distribution. For such a case, a portion of the ET region is not ruled out by the current constraints, while the LISA region is almost completely excluded. 

For the case of clustered PBHs, the comparison between the forecast and the current constraints is more delicate, given that some of them might be altered by PBH clustering. However, we point out that the most stringent constraints, such as the ones from CMB and X-ray observations, are not weakened if PBHs are clustered~\cite{ Manshanden:2018tze,Hutsi:2019hlw,DeLuca:2020jug}. Finally, we remark that constraints applying on early universe quantities, like CMB anisotropies mostly sensitive to the epoch $300\lesssim z\lesssim 600$, may be potentially evaded if accretion is strong enough to shift them to larger final masses~\cite{DeLuca:2020fpg}. This effect can impact on constraints at masses $m_\PBH \gtrsim {\cal O}(10) M_\odot$ and may be of particular interest for LISA.

The forecast for the most conservative minimum value of $f_\PBH$ described in Fig.~\ref{fig: 3} is obtained supposing that PBHs will provide a signal seen by future detectors (i.e. $N_\text{\tiny det}\gtrsim 1/{\rm  yr}$).
While a PBH population above the minimum $f_\PBH$ would contribute to the observed events, one would not necessarily be able to differentiate it from other astrophysical contributions, unless one focuses on subsolar mergers (see Refs.~\cite{Abbott:2018oah,
Authors:2019qbw,
Wang:2021qsu,
Nitz:2021mzz,
Nitz:2021vqh} for constraints with current data). 
In this sense, the bound discussed in this section can be regarded as a necessary, but not sufficient, condition allowing for a test of the existence of PBHs due to the difficulty of distinguishing the signal from  the background of astrophysical mergers. 
In Sec.~\ref{sec2}, we will require a tighter condition defining the parameter space where a detection operated by future detectors can be confidently ascribed to PBHs, i.e. restricting to high enough redshifts where there is no astrophysical contamination.

\subsection{Stochastic gravitational wave background}
\begin{figure}[t!]
	\centering
		\includegraphics[width=0.49 \linewidth]{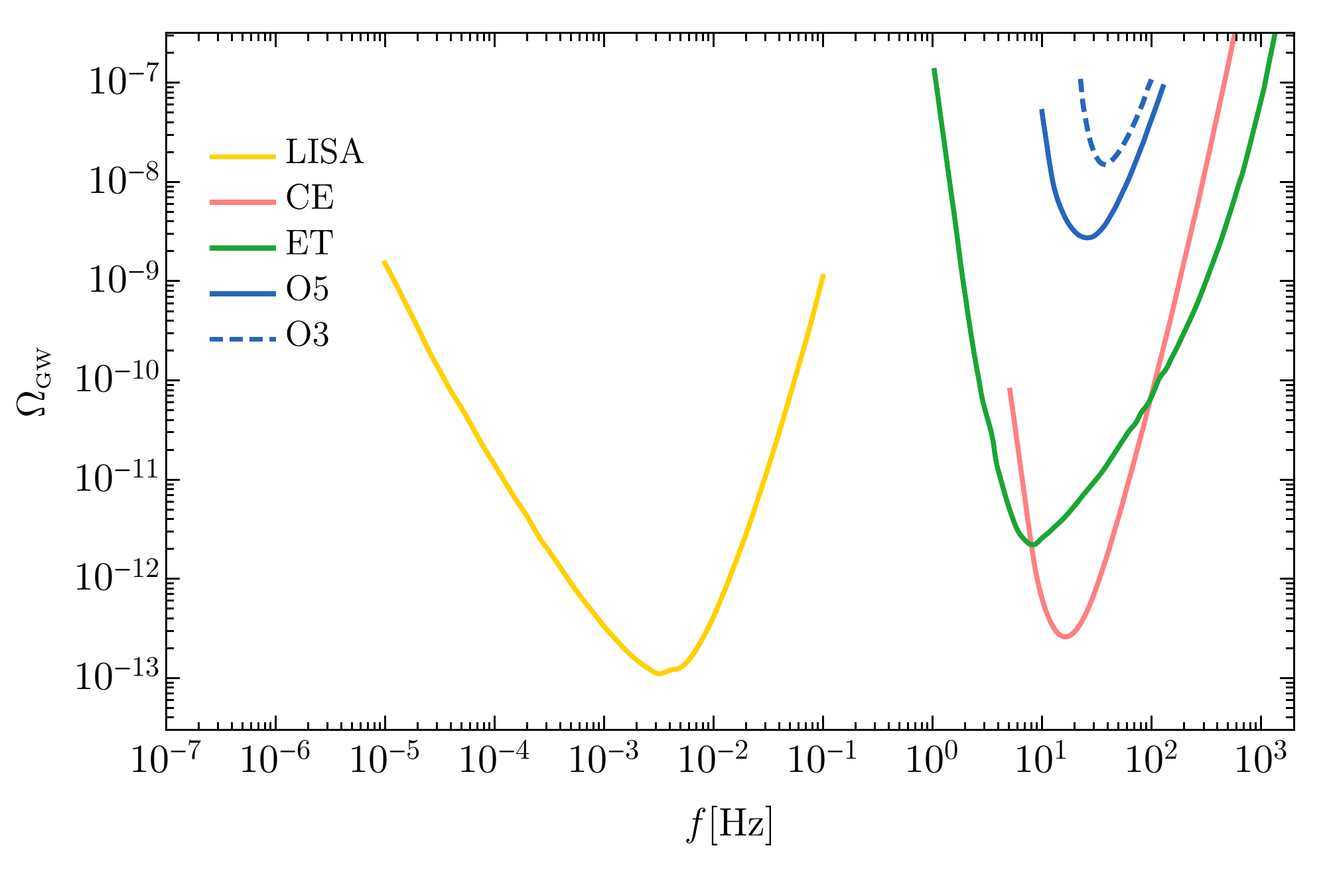}
	\includegraphics[width=0.49 \linewidth]{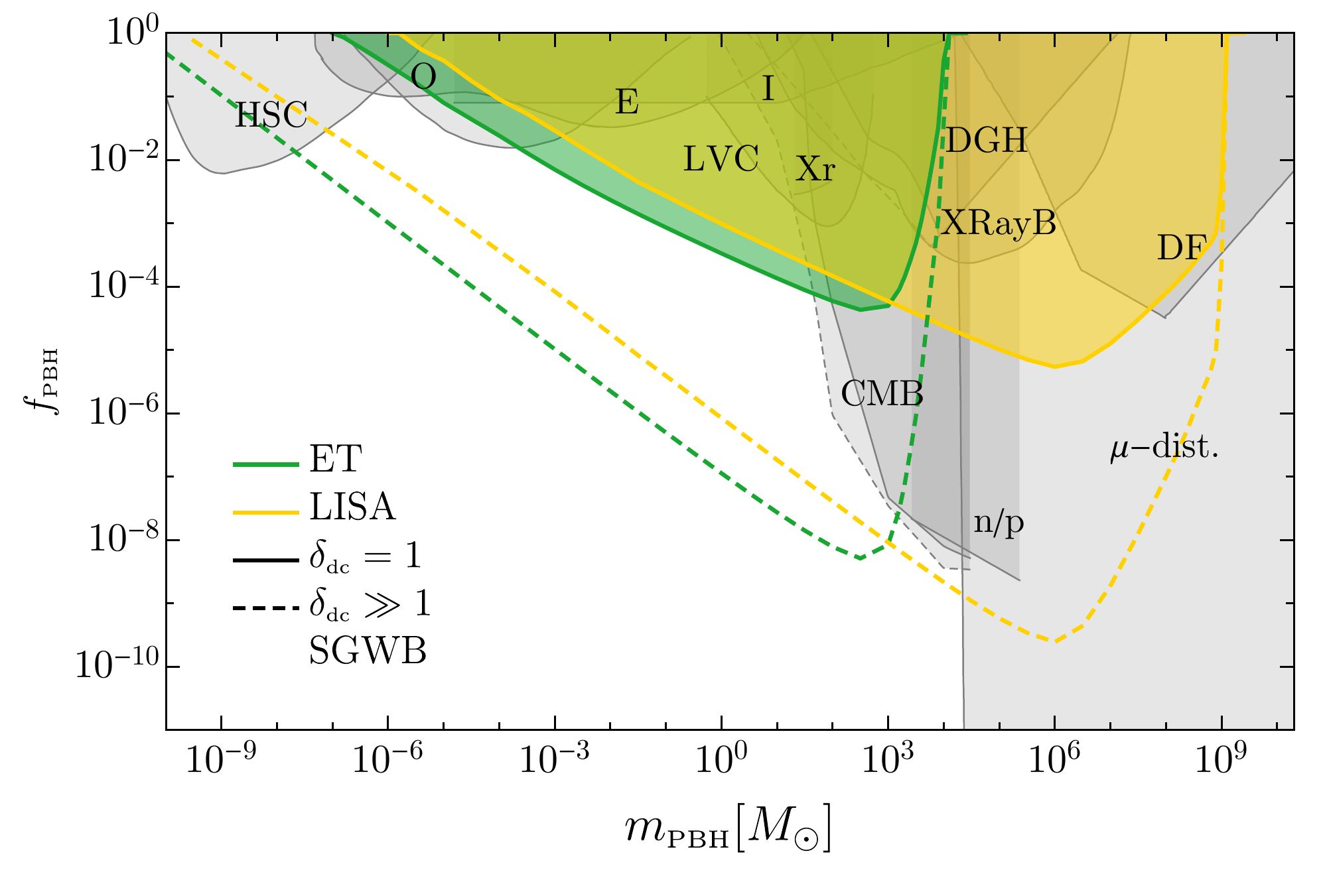}
	\caption{\it
	{\bf Left:} Sensitivity curves for the SGWB for several present and future GW experiments.
	{\bf Right:}
	Forecast on the PBH abundance coming from the SGWB at the ET (green) and LISA (yellow) experiments, for both cases of Poisson distributed (solid) and clustered (dashed) PBHs. As done in Fig. 3, we superimpose the constraints coming from independent observations.}
	\label{fig: 4}
\end{figure}

\noindent
PBH mergers which are not individually resolved may give rise to a Stochastic Gravitational Wave Background (SGWB), whose spectrum at frequency $\nu$ is given by
\be
\Omega_\text{\tiny GW} (\nu)=  \frac{\nu}{\rho_0} \int_0^{\frac{\nu_3}{\nu}-1} \d z \d m_1 \d m_2 \,\frac{1}{(1+z)H(z)}  \frac{\d R_\PBH}{\d m_1 \d m_2}  \frac{\d E_\text{\tiny GW} (\nu_s)}{\d \nu_s},
\ee
in terms of the redshifted source frequency $\nu_s =\nu (1+z)$, the present energy density $\rho_0 = 3 H_0^2/8\pi$ in terms of the Hubble constant $H_0$, and the energy spectrum of GWs. The latter is given by the phenomenological expression in the non-spinning limit~\cite{Ajith:2009bn} 
\be
\frac{\d E_\text{\tiny GW} (\nu)}{\d \nu} =  \frac{\pi^{2/3}}{3} M_\text{\tiny tot}^{5/3} \eta \times
\left \{ \begin{array}{rl}
	&\nu^{-1/3} \lp 1+ \alpha_2 \nu^2\rp^2 \qquad \qquad \quad  \text{for} \quad \nu < \nu_1,  \\
	& w_1 \nu^{2/3} \lp 1+ \epsilon_1 \nu + \epsilon_2 \nu^2 \rp^2 \qquad  \text{for} \quad \nu_1 \leq \nu < \nu_2, \\
	& w_2 \nu^2 \frac{ \sigma^4}{(4 (\nu- \nu_2)^2 + \sigma^2)^2} \quad \quad \quad \quad \ \ \text{for}  \quad \nu_2 \leq \nu < \nu_3 , 
\end{array}
\right.
\ee
where $\eta = m_1 m_2/M_\text{\tiny tot}^2$, $\alpha_2 = -{323}/{224} + \eta\, {451}/{168}$, $\epsilon_1 = -1.8897$, $\epsilon_2 = 1.6557$,
\begin{align}
	&w_1  = \nu_1^{-1} \frac{[1 + \alpha_2 (\pi M_\text{\tiny tot} \nu_1)^{2/3}]^2}{[1+ \epsilon_1 (\pi M_\text{\tiny tot} \nu_1)^{1/3} +\epsilon_2 (\pi M_\text{\tiny tot} \nu_1)^{2/3}]^2},  \nonumber \\
	&w_2  = w_1 \nu_2^{-4/3} [1+ \epsilon_1 (\pi M_\text{\tiny tot} \nu_2)^{1/3} +\epsilon_2 (\pi M_\text{\tiny tot} \nu_2)^{2/3}]^2,  
\end{align}
and
\begin{align}
	&\pi M_\text{\tiny tot} \nu_1  = (1-4.455+3.521)+0.6437\eta-0.05822\eta^2-7.092\eta^3,  \nonumber \\
	&\pi M_\text{\tiny tot} \nu_2  = (1-0.63)/2+0.1469\eta-0.0249\eta^2+2.325\eta^3,  \nonumber \\
	&\pi M_\text{\tiny tot} \sigma = (1-0.63)/4 -0.4098\eta +1.829\eta^2-2.87\eta^3,  \nonumber \\
	& \pi M_\text{\tiny tot} \nu_3 = 0.3236 -0.1331\eta -0.2714\eta^2 +4.922\eta^3.
\end{align}
The comparison of the strength of the SGWB with the sensitivity curves of various GW experiments (see the left panel of Fig.~\ref{fig: 4} for a plot of those curves)
can be used to make a forecast for the minimum  PBH abundance in the universe, see for example Refs.~\cite{Wang:2016ana, Raidal:2017mfl, Chen:2018rzo, Raidal:2018bbj, DeLuca:2020qqa}.
The result is shown in the right panel of Fig.~\ref{fig: 4} for the ET and LISA experiments, for both cases of Poisson distributed and clustered PBHs.

One can notice first that, given the enhancement of the merger rate for strong clustering, the forecast  becomes much more stringent in this case with respect to the Poisson case. Furthermore, by comparing this result with the one obtained in Fig.~\ref{fig: 3}, one can appreciate that the forecast coming from the SGWB seems to be less stringent that the one obtained from the number of events for the ET experiment, even though the former applies to a broader range of PBH masses reaching smaller values. For LISA the constraint reaches similar values for the PBH abundance, even though it extends to much smaller masses. The reason is that lighter PBH populations may still cross the LISA detectability band due to the low frequency tail scaling as $\sim f^{2/3}$ characterizing the SGWB.

\section{Minimum abundance for PBH evidence}
\renewcommand{\theequation}{3.\arabic{equation}}
\setcounter{equation}{0}
\label{sec2}
\noindent
The characteristic evolution with redshift of the PBH merger rate, for both cases of Poisson and clustered spatial distribution, is very different with respect to the merger rate evolution of astrophysical BHs. In particular, the PBH behaviour is found to be monotonically increasing in redshift following the power laws $R_\text{\tiny PBH} \sim (t/t_0)^{-34/37}$ for the Poisson case and $R_\text{\tiny PBH} \sim (t/t_0)^{-1}$ for the clustered one. By contrast, the merger rate of astrophysical populations is expected to have a peak around redshift of a few, with a possible second peak due to the merger of BHs from Pop III stars~\cite{Schneider:1999us,Schneider:2001bu,Schneider:2003em,Liu:2020ufc} at redshift $z\sim 10$~\cite{Ng:2020qpk,Kinugawa:2014zha,Kinugawa:2015nla,Hartwig:2016nde}. Those BHs are expected to form at $z\sim 25$~\cite{Valiante:2020zhj} and their time of merger mainly depends on the binary formation mechanisms, and could vary from ${\cal O}({\rm Gyr})$ (resulting in 
merger at $z\lesssim6$) to ${\cal O}(10\,{\rm Myr})$ if they form in Pop~III clusters with dynamical mechanisms, in which case they would merge almost at the redshift of BH formation~\cite{Valiante:2020zhj}.

Assuming the conservative scenario in which astrophysical BH coalescences are formed from Pop~III clusters up to redshift  $z=30$, the detection of a binary system at larger redshifts at future experiments with larger horizons like the ET or LISA  would be a smoking-gun in favour of PBHs, given that no astrophysical BHs are expected to form in standard cosmologies at  those or higher redshifts~\cite{Koushiappas:2017kqm}. Similarly a smoking gun for PBH evidence would be a detection of subsolar PBH masses, where the results of Fig.~\ref{fig: 3} 
are sufficient.

\begin{figure}[t!]
	\centering
	\includegraphics[width=0.6 \linewidth]{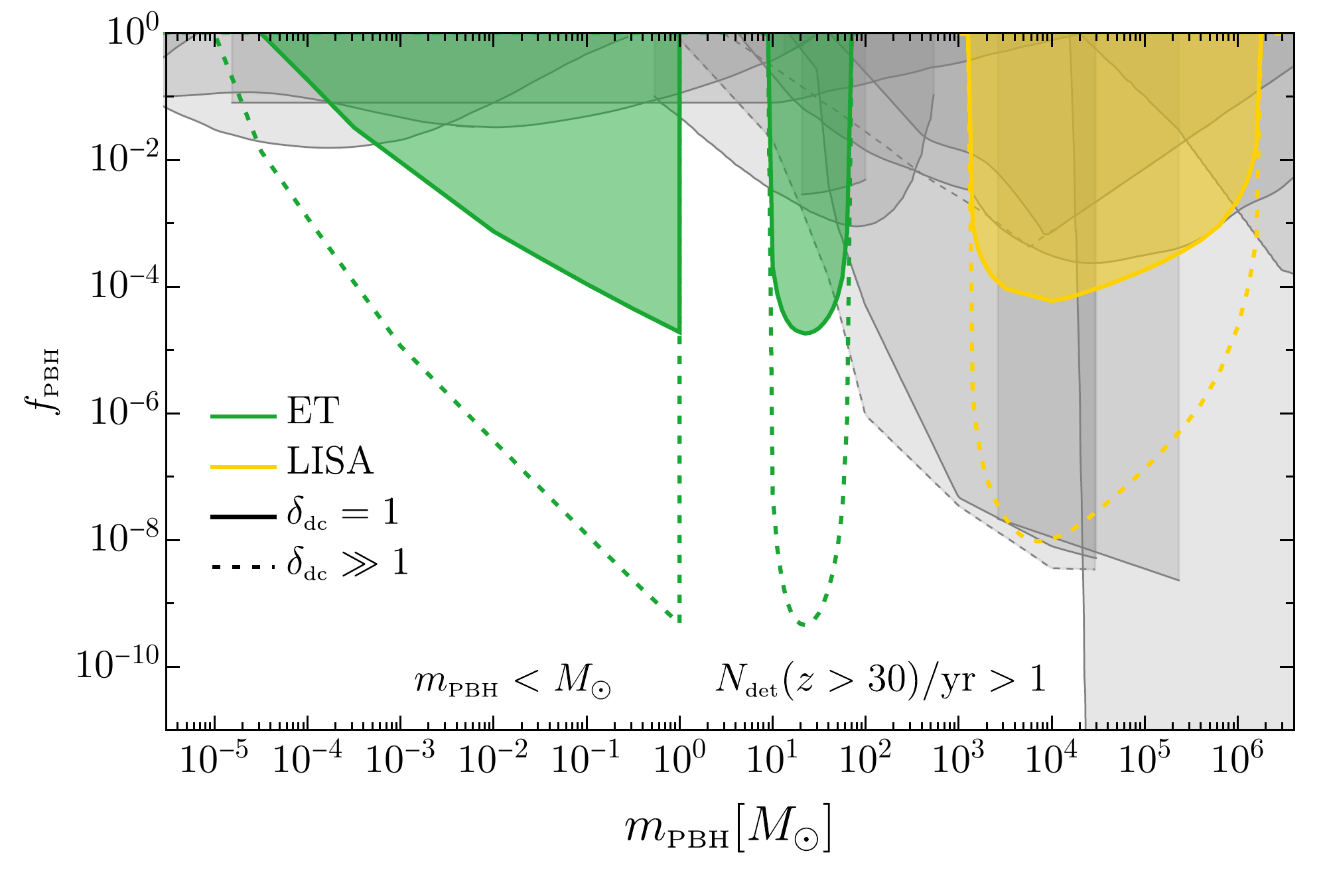}
	\caption{\it Minimum  abundance for PBH evidence:  above the Chandrasekhar's limit around $M_\odot$ we require  at least an observable event per year at redshift larger than 30;  in the mass range below  the Chandrasekhar's limit we apply the same requirement of section III, see Fig. 3, i.e. the integrated (in redshifts) number of events per year to be larger than unity.}
	\label{fig: 5}
\end{figure}

Following the same logic of the previous section, one can therefore estimate the minimum PBH abundance required to give at least one event per year at ET and LISA considering redshifts larger than 30, in order to get an evidence of PBH mergers.\footnote{A more refined approach may be performing a population analysis of the merger rate evolution at high redshift to distinguish PopIII and PBH populations~\cite{inprep}. } The result is shown in Fig.~\ref{fig: 5} for both cases of Poisson and clustered spatial distributions.
One can notice that, even though the minimum value of $f_\PBH$ does not strongly differ with respect to the case of integrating over all redshifts as in Fig.~\ref{fig: 3}, the constraint applies to a smaller range of masses for both the experiments. The reason lies on the characteristic shapes of the horizon redshift for the two experiments shown in Fig.~\ref{fig: 3}, which become narrower at larger redshifts, limiting therefore the range of masses which could be possibly detected. This effect is much stronger for ET than LISA, however, the contribution coming from large masses is strongly suppressed in both cases.

\section{Conclusions} 
\renewcommand{\theequation}{4.\arabic{equation}}
\setcounter{equation}{0}
\label{sec:conclusions}

\noindent 
PBHs may represent a good candidate for the dark matter in the universe as they would not require any exotic new physics beyond the Standard Model. 
In this paper we have addressed the question of what is  the minimum testable contribution of PBHs to the dark matter of the universe by future detectors.
Our findings indicate that an abundance as small as $10^{-10}$ can be probed, in different mass ranges, by both 3G detectors and LISA in the case in which PBHs are clustered. Similar figures are obtained by restricting to sufficiently large redshifts, even though in a smaller PBH mass range. We stress again that going to large redshifts is not necessary when dealing with subsolar masses, for which Fig.~\ref{fig: 3} provides the relevant forecast.

As a final remark, we notice that the current constraints already eliminate portions of the 3G forecast regions. We stress that such constraints are valid within a given set of hypotheses, such as monochromatic PBH mass functions, no PBH clustering and no accretion effects. It would be interesting to investigate the impact of relaxing these assumptions on our results.

\vskip 0.3cm
\noindent
\section*{Acknowledgments}
\noindent
We acknowledge use of the software package {\tt pycbc}~\cite{alex_nitz_2021_4556907} and, for the computation of the SNR of mergers at LISA, {\tt gwent} \cite{Kaiser:2020tlg}.
V.DL., G.F. and  A.R. are supported by the Swiss National Science Foundation (SNSF), project {\sl The Non-Gaussian Universe and Cosmological Symmetries}, project number: 200020-178787.
P.P. acknowledges financial support provided under the European Union's H2020 ERC, Starting 
Grant agreement no.~DarkGRA--757480, and under the MIUR PRIN and FARE programmes (GW-NEXT, CUP:~B84I20000100001), and 
support from the Amaldi Research Center funded by the MIUR program `Dipartimento di 
Eccellenza" (CUP:~B81I18001170001).

\bigskip

\bibliography{main}
 
\end{document}